\documentclass[%
 reprint,
superscriptaddress,
 amsmath,amssymb,
 aps,
 prl,
]{revtex4-1}

\usepackage{graphicx}

\usepackage{siunitx}
\usepackage{ulem}

\usepackage{xcolor}

\usepackage{subcaption}
\usepackage[utf8x]{inputenc}

\usepackage{supertabular}
\usepackage{hyperref}

\makeatletter
\renewcommand\@makecaption[2]{%
  \par
  \vskip\abovecaptionskip
  \begingroup
   \small\rmfamily
    \begingroup
     \samepage
     \flushing
     \let\footnote\@footnotemark@gobble
     \@make@capt@title{#1}{#2}\par
    \endgroup
  \endgroup
  \vskip\belowcaptionskip
}
\makeatother
\begin{document}
\title{Dark Matter Search Results from the PandaX-4T Commissioning Run}


\def\shKeyLab{School of Physics and Astronomy, Shanghai Jiao Tong University, MOE Key Laboratory for Particle Astrophysics and Cosmology, Shanghai Key Laboratory for Particle Physics and Cosmology, Shanghai 200240, China}
\def\BUAA{School of Physics, Beihang University, Beijing 100191, China}
\def\USTClab{State Key Laboratory of Particle Detection and Electronics, University of Science and Technology of China, Hefei 230026, China}
\def\USTCdep{Department of Modern Physics, University of Science and Technology of China, Hefei 230026, China}
\def\BUAALab{International Research Center for Nuclei and Particles in the Cosmos \& Beijing Key Laboratory of Advanced Nuclear Materials and Physics, Beihang University, Beijing 100191, China}
\def\pku{School of Physics, Peking University, Beijing 100871, China}
\def\YaLongSD{Yalong River Hydropower Development Company, Ltd., 288 Shuanglin Road, Chengdu 610051, China}
\def\IAP{Shanghai Institute of Applied Physics, Chinese Academy of Sciences, 201800 Shanghai, China}
\def\CHEPpku{Center for High Energy Physics, Peking University, Beijing 100871, China}
\def\SDUdep{Research Center for Particle Science and Technology, Institute of Frontier and Interdisciplinary Science, Shandong University, Qingdao 266237, Shandong, China}
\def\SDUlab{Key Laboratory of Particle Physics and Particle Irradiation of Ministry of Education, Shandong University, Qingdao 266237, Shandong, China}
\def\UMD{Department of Physics, University of Maryland, College Park, Maryland 20742, USA}
\def\TDLee{Tsung-Dao Lee Institute, Shanghai Jiao Tong University, Shanghai, 200240, China}
\def\MESJTU{School of Mechanical Engineering, Shanghai Jiao Tong University, Shanghai 200240, China}
\def\SYU{School of Physics, Sun Yat-Sen University, Guangzhou 510275, China}
\def\NKU{School of Physics, Nankai University, Tianjin 300071, China}
\def\FDU{Key Laboratory of Nuclear Physics and Ion-beam Application (MOE), Institute of Modern Physics, Fudan University, Shanghai 200433, China}
\def\USST{School of Medical Instrument and Food Engineering, University of Shanghai for Science and Technology, Shanghai 200093, China}
\def\SJTUSC{Shanghai Jiao Tong University Sichuan Research Institute, Chengdu 610213, China}
\def\Princeton{Physics Department, Princeton University, Princeton, NJ 08544, USA}
\def\MIT{Department of Physics, Massachusetts Institute of Technology, Cambridge, MA 02139, USA}
\def\SARI{Shanghai Advanced Research Institute, Chinese Academy of Sciences, Shanghai 201210, China}
\def\SPEIT{SJTU Paris Elite Institute of Technology, Shanghai Jiao Tong University, Shanghai, 200240, China}

\author{Yue Meng}\affiliation{\shKeyLab}\affiliation{\SJTUSC}
\author{Zhou Wang}\affiliation{\shKeyLab}\affiliation{\SJTUSC}\affiliation{\TDLee}
\author{Yi Tao}\affiliation{\shKeyLab}\affiliation{\SJTUSC}
\author{Abdusalam Abdukerim}
\author{Zihao Bo}
\author{Wei Chen}\affiliation{\shKeyLab}
\author{Xun Chen}\affiliation{\shKeyLab}\affiliation{\SJTUSC}
\author{Yunhua Chen}\affiliation{\YaLongSD}
\author{Chen Cheng}\affiliation{\SYU}
\author{Yunshan Cheng}\affiliation{\SDUdep}\affiliation{\SDUlab}
\author{Xiangyi Cui}\affiliation{\TDLee}
\author{Yingjie Fan}\affiliation{\NKU}
\author{Deqing Fang}
\author{Changbo Fu}\affiliation{\FDU}
\author{Mengting Fu}\affiliation{\pku}
\author{Lisheng Geng}\affiliation{\BUAA}\affiliation{\BUAALab}
\author{Karl Giboni}
\author{Linhui Gu}\affiliation{\shKeyLab}
\author{Xuyuan Guo}\affiliation{\YaLongSD}
\author{Ke Han}\affiliation{\shKeyLab}
\author{Changda He}\affiliation{\shKeyLab}
\author{Jinrong He}\affiliation{\YaLongSD}
\author{Di Huang}\affiliation{\shKeyLab}
\author{Yanlin Huang}\affiliation{\USST}
\author{Zhou Huang}\affiliation{\shKeyLab}
\author{Ruquan Hou}\affiliation{\SJTUSC}
\author{Xiangdong Ji}\affiliation{\UMD}
\author{Yonglin Ju}\affiliation{\MESJTU}
\author{Chenxiang Li}\affiliation{\shKeyLab}
\author{Mingchuan Li}\affiliation{\YaLongSD}
\author{Shu Li}\affiliation{\MESJTU}
\author{Shuaijie Li}\affiliation{\TDLee}
\author{Qing  Lin}\email[Corresponding author: ]{qinglin@ustc.edu.cn}\affiliation{\USTClab}\affiliation{\USTCdep}
\author{Jianglai Liu}\email[Spokesperson: ]{jianglai.liu@sjtu.edu.cn}\affiliation{\shKeyLab}\affiliation{\TDLee}\affiliation{\SJTUSC}
\author{Xiaoying Lu}\affiliation{\SDUdep}\affiliation{\SDUlab}
\author{Lingyin Luo}\affiliation{\pku}
\author{Wenbo Ma}\affiliation{\shKeyLab}
\author{Yugang Ma}\affiliation{\FDU}
\author{Yajun Mao}\affiliation{\pku}
\author{Nasir Shaheed}\affiliation{\SDUdep}\affiliation{\SDUlab}
\author{Xuyang Ning}\affiliation{\shKeyLab}
\author{Ningchun Qi}\affiliation{\YaLongSD}
\author{Zhicheng Qian}\affiliation{\shKeyLab}
\author{Xiangxiang Ren}\affiliation{\SDUdep}\affiliation{\SDUlab}
\author{Changsong Shang}\affiliation{\YaLongSD}
\author{Guofang Shen}\affiliation{\BUAA}
\author{Lin Si}\affiliation{\shKeyLab}
\author{Wenliang Sun}\affiliation{\YaLongSD}
\author{Andi Tan}\affiliation{\UMD}
\author{Anqing Wang}\affiliation{\SDUdep}\affiliation{\SDUlab}
\author{Meng Wang}\affiliation{\SDUdep}\affiliation{\SDUlab}
\author{Qiuhong Wang}\affiliation{\FDU}
\author{Shaobo Wang}\affiliation{\shKeyLab}\affiliation{\SPEIT}
\author{Siguang Wang}\affiliation{\pku}
\author{Wei Wang}\affiliation{\SYU}
\author{Xiuli Wang}\affiliation{\MESJTU}
\author{Mengmeng Wu}\affiliation{\SYU}
\author{Weihao Wu}
\author{Jingkai Xia}\affiliation{\shKeyLab}
\author{Mengjiao Xiao}\affiliation{\UMD}
\author{Xiang Xiao}\affiliation{\SYU}
\author{Pengwei Xie}\affiliation{\TDLee}
\author{Binbin Yan}\affiliation{\shKeyLab}
\author{Xiyu Yan}\affiliation{\USST}
\author{Jijun Yang}
\author{Yong Yang}\affiliation{\shKeyLab}
\author{Chunxu Yu}\affiliation{\NKU}
\author{Jumin Yuan}\affiliation{\SDUdep}\affiliation{\SDUlab}
\author{Ying Yuan}\affiliation{\shKeyLab}
\author{Dan Zhang}\affiliation{\UMD}
\author{Minzhen Zhang}\affiliation{\shKeyLab}
\author{Peng Zhang}\affiliation{\YaLongSD}
\author{Tao Zhang}
\author{Li Zhao}\affiliation{\shKeyLab}
\author{Qibin Zheng}\affiliation{\USST}
\author{Jifang Zhou}\affiliation{\YaLongSD}
\author{Ning Zhou}\email[Corresponding author: ]{nzhou@sjtu.edu.cn}\affiliation{\shKeyLab}
\author{Xiaopeng Zhou}\email[Corresponding author: ]{zhou\_xp@buaa.edu.cn}\affiliation{\BUAA}
\author{Yong Zhou}\affiliation{\YaLongSD}

\collaboration{PandaX-4T Collaboration}
\noaffiliation

\date{\today}

\begin{abstract}
We report the first dark matter search results using the commissioning data from PandaX-4T. Using a time projection chamber with 3.7-tonne of liquid xenon target and an exposure of 0.63~tonne$\cdot$year, 1058 candidate events are identified within an approximate nuclear recoil energy window between 5 and 100 keV.
No significant excess over background is observed. Our data set a stringent limit to the dark matter-nucleon spin-independent interactions, with a lowest excluded cross section (90\% C.L.) of $3.8\times10^{-47} $cm$^2$ at a dark matter mass of 40~GeV/$c^2$. 
\end{abstract}

\maketitle
Like ordinary matter, the mysterious dark matter in the Universe may be composed of fundamental particles~\cite{BERTONE2005279review}. The hunt for these particles has been intensively carried out globally using many different particle detectors~\cite{NPreviewDirect,NPreviewIndirect,NPreviewCollider,APPECreport}. Dark matter direct detection experiments, typically located deep underground, are particularly sensitive to dark matter within a mass range approximately from GeV/$c^2$ to 100 TeV/$c^2$, via the nuclear recoil (NR) of the target nucleus~\cite{NPreviewDirect}. In recent years, 
large-scale liquid xenon time projection chambers (TPCs) have spearheaded the detection sensitivity~\cite{Akerib:2016vxi_LUXNR,Cui:2017nnnPandaXII,Aprile:2018dbl_xenon1tNR,finalcpc}, and three new experiments with multitonne of targets are ongoing to deepen the search~\cite{hongguang,Mount:2017qzi,XENON:2020kmp}.

The PandaX experiment, located in the China Jinping Underground Laboratory (CJPL)~\cite{Kang:2010zza}, is dedicated to search for the dark matter particles and to study fundamental properties of neutrinos. PandaX-4T~\cite{hongguang}, with a sensitive target of 3.7 tonne of liquid xenon, is located in the B2 hall of the newly expanded CJPL-II~\cite{Li:2014rca}. The detector is placed at the center of an ultrapure water shield in a stainless steel tank with a diameter of 10 m and a depth of 13 m. The double-vessel cryostat made out of low-background stainless steel~\cite{Zhang:2016pgh} contains 5.6 tonne of total liquid xenon, with a 30-l overflow chamber inside the cryostat for adjusting the liquid level~\cite{cao2014pandax}. A cryogenic system containing three independent cold heads is constantly delivering cooling power (580\,W at maximum) to liquefy xenon~\cite{Zhao:2020vxh}. 
The xenon is being continuously purified by two hot metal getters manufactured by $SAES$~\footnote{SAES Pure Gas. \url{https://www.entegris.com/shop/en/USD/Products/Gas-Filtration-and-Purification/Gas-Purifiers/c/gaspurifiers}, 2021 (accessed November 30, 2021).}, through two separate circulation loops with stable flow rates of about 80 and 30 standard-liter-per-minute (slpm), respectively. Purified xenon gas is driven by diaphragm pumps into two heat exchangers located close to the detector. Inside the heat exchangers, purified gas is cooled and liquefied by liquid xenon extracted from the detector.

The sensitive target is a cylindrical dual phase xenon TPC confined by 24 highly reflective polytetrafluoroethylene (PTFE) wall panels, with an opposite-panel distance of 1185\,mm (room temperature). The electrical fields in the TPC are defined by, from the bottom to the top, a cathode grid, a gate mesh, and an anode mesh, with a separation of 1185\,mm and 10\,mm in between. The liquid level is set in between the gate and anode by the top opening of the overflow tube, which is adjustable externally via a motion feedthrough. 
Under an electrical field, the average gas gap is 3.5\,mm from the anode, and the relative distortion between the gate and anode, primarily due to electrostatic attraction, is less than 0.4\,mm. 
A total of 169 and 199 of Hamamatsu R11410-23 three-inch photomultipliers (PMTs) are located at the top and bottom of the TPC, respectively, with grounded screening meshes 6\,mm away from the PMT surfaces. 
During the operation, nine R11410-23 PMTs were turned off due to connection or base problems, and four PMTs were turned off due to excessive noise. The average dark rate for the remaining PMTs is about 100\,Hz per channel.
The prompt scintillation photons ($S1$), and delayed electroluminescence photons ($S2$, proportional to the number of ionized electrons extracted into the gaseous region) are measured by the top and bottom PMT arrays. This allows, for a given event, a three-dimensional vertex reconstruction to a subcentimeter precision.
The outside wall of the field cage is about 70\,mm from the inner cryostat, to leave enough space for the cathode feedthrough. Two rings of Hamamatsu R8520 one-inch PMTs (105 in total) are instrumented in this gap facing upward and downward, respectively, serving as the background veto. 
The PMT gains are calibrated, once per week, by four external blue light-emitting diodes with photons transmitted into the detector via
optical fibers. The average gains of the PMTs are $5.5\times10^{6}$ for R11410-23 and $2.3\times10^{6}$ for R8520.
The PMT pulses are amplified by low-noise linear amplifiers with a gain of 1.5 and 5 for R11410-23 and R8520 PMTs, respectively, and then digitized by the CAEN V1725B digitizers with 0.122~mV per analog-to-digital-convertor(ADC) bit and a sampling rate of $2.5\times10^8$ samples per second~\footnote{see \url{https://www.caen.it/products/v1725/}.}. The digitizers are operated under the self-trigger mode, so if any pulse is above a predefined threshold corresponding to about 1/3 of a photoelectron (PE), the entire waveform is read out~\cite{Zheng:2020kfp, Yang:2108}. The readout efficiency for a single PE is measured for each channel, with an average value of 96\%.
 The data are read out through optical fibers and directly stored onto the disk. Physical events are reconstructed via an off-line software~\footnote{X. Chen $et~al.$, in preparation.}.

An off-line krypton distillation was carried out on all 5.6 tonne of xenon using a newly constructed distillation tower at CJPL~\cite{Cui:2020bwf}. The detector was then filled, and after basic functionality checks, the water shield was 
filled with ultrapure water which has electrical resistivity of about 18\,MOhms$\cdot$cm and concentration of uranium and thorium less than 0.1\,ppt. The commissioning run of PandaX-4T commenced on November \,28, 2020 and ended on Aprile\,16, 2021, including 95.0 calendar days of stable data taking. In this period, the diaphragms of the circulation pumps were worn out two times, each time causing degradation in electron drifting, but with no trace of radioactive impurity introduced. The cathode and gate voltages were set at several different values to avoid excessive discharges, separating the data into several sets. The liquid level was adjusted between sets 2 and 3. 
During set 4, the online krypton distillation was kept on with a flow rate of 10~slpm. The detailed run configurations can be found in Table~\ref{table:runConfig}.  

\begin{table}[htb]
\caption{\label{table:runConfig}Basic detector configurations of the commissioning datasets, with $\langle\tau_e\rangle$, $dt_{\rm{max}}$, PDE, EEE, and $\rm SEG_{b}$ representing the average electron lifetime, maximum drift time, photon detection efficiency, electron extraction efficiency, and single-electron gain from the bottom PMT array (with a total-to-bottom ratio of 4.2), respectively.}
\begin{tabular}{lccccc}
\\\hline\hline
    Set   & 1     & 2     & 3     & 4     & 5 \\ \hline
    Duration (days)  & 1.95  & 13.25 & 5.53  & 35.58 & 36.51 \\
    $\langle\tau_e\rangle$ ($\mu$s) & 800.4 & 939.2 & 833.6 & 1121.5 & 1288.2 \\
    $dt_{\rm{max}}$ ($\mu$s) & 800 & 810 & 817 & 841 & 841 \\
    $V_{\rm cathode}$ ($-$kV) & 20 & 18.6 & 18 & 16 & 16 \\
    $V_{\rm gate}$ ($-$kV) & 4.9 & 4.9 & 5 & 5 & 5 \\
    \hline
    PDE (\%)  & \multicolumn{2}{c|}{9.0$\pm$0.2} & \multicolumn{3}{c}{9.0$\pm$0.2}\\
    EEE (\%)  & \multicolumn{2}{c|}{90.2$\pm$5.4} & \multicolumn{3}{c}{92.6$\pm$5.4} \\
    $\rm SEG_{b}$ (PE/$e$)  & \multicolumn{2}{c|}{3.8$\pm$0.1} & \multicolumn{3}{c}{4.6$\pm$0.1} \\\hline\hline
\end{tabular}
\end{table}

The data processing follows a similar procedure as in the previous PandaX analysis~\cite{finalcpc}. Hits with amplitudes larger than 20\,ADC ($\sim$2.44\,mV) are identified from the waveform of individual channels. Signals are defined as clusters of hits with a tail-to-head gap no greater than 15 samples (60\,ns), corresponding to an approximate 104\,ns peak-to-peak separation between hits, and a coincidence requirement that at least two different PMTs receive hits. The inefficiency of such clustering gap requirement is validated to be negligible using data-driven approaches.
Other unphysical noises are identified by anomalous shapes or charge distribution pattern. Signals are then classified into $S1$-like and $S2$-like according to number of hits, the charge ratio between the top and bottom arrays, and width of the waveform enclosing $10\%-90\%$ cumulative charge ($w_{\rm cum}$). 
$S2$-like signals are reclustered by taking into account the diffusion effect during the drift. The inefficiency of tagging is verified to be negligible. $S1$-like and $S2$-like signals within a window of 1~ms are further combined into an event.

Three classes of data quality cuts are developed based on the calibration data (see later) and on our practice in previous generations of PandaX, to remove noise and unphysical events.
(i) A set of ``waveform cleanliness" cuts is applied to avoid having too much noise or too many single-electron $S2$s within an event. (ii) For $S1$s, to avoid confusions with single-electron $S2$s, the number of peaks in a summed waveform is required to be no more than four, and the top-bottom charge ratio should be consistent with the location of the interaction. Another cut is applied on the charge distribution to suppress abnormal charge caused by PMT after pulsing.
(iii) For $S2$s, cuts are applied to their waveform shapes, top-bottom charge ratio, the root-mean-square (rms) of the 
charge distribution on the top PMTs, and the quality of horizontal position reconstruction.
An $S2$-dependent cut on drift time vs. $w_{\rm{cum}}$ is made, which is important to suppress the accidental background. Unless otherwise specified, a good event should have only one pair of $S1$ and $S2$.

The position of an event is reconstructed using the charge pattern on the top PMT array (horizontal) and the drift time (vertical) assuming a constant drift velocity. Two independent horizontal reconstruction methods have been developed,
the template matching method (TM) and the photon acceptance function method (PAF)~\cite{PANDA-X:2021jua}. 
Only the first half of the charge in $S2$ is used in the reconstruction, leading to a better position resolution in comparison to that using the total charge. The position uncertainty in the vertical direction is conservatively estimated to be 3\,mm, based on the width of the $S2$. In the horizontal plane, the reconstruction uncertainty depends on the charge of $S2$, and is estimated to be 8.2\,mm (100\,PE) and 3.0\,mm (1000\,PE) based on the comparison between the two methods, cross-checked with the sigma of the radial distribution of the surface events from the PTFE wall.


The uniformity of detector responses is calibrated using internal diffusive sources. To avoid PMT saturations, the bottom-only $S2$ ($S2_{\rm b}$) is used. The vertical uniformity of $S2_{\rm b}$, characterized by the electron lifetime $\tau_{\rm e}$, is calibrated using the 164 keV deexcitation peak from $^{131\rm m}$Xe (produced by neutron irradiation), which is found consistent with that obtained from radon alpha peaks. The average $\tau_{\rm e}$ in each run set is summarized in Table~\ref{table:runConfig}.  The three-dimensional uniformity of $S1$ and the horizontal uniformity of $S2_{\rm b}$ are calibrated by injecting $^{83\rm m}$Kr (41.5 keV) into the detector via one of the circulation loops~\cite{Zhang:2021shp}. This calibration was carried out twice, at the beginning and completion of the commissioning data taking, with about 100,000 events collected. In the fiducial volume (FV, defined later), the rms variation in $S1$ and horizontal $S2_{\rm b}$ responses is 19\% and 15\%, respectively. 

The electron-equivalent energy $E$ of a given event can be reconstructed as~\cite{Lenardo_2015}
\begin{equation}
    E = 13.7~\rm{eV} \times (\frac{S1}{\rm{PDE}} + \frac{S2_{\rm b}}{\rm{EEE\times SEG_{\rm b}}})\,,
\label{eq:doke}
\end{equation}
in which PDE, EEE, and SEG$_{\rm b}$ are the photon detection efficiency for $S1$, electron extraction efficiency, and the single-electron gain using $S2_{\rm b}$, respectively, and 13.7\,eV is the work function in LXe. The value of SEG$_{\rm b}$ is measured by selecting the smallest $S2_{\rm b}$, with an enlarged 50-sample clustering gap requirement. 
The rms of the SEG$_{\rm b}$ in the FV is 8\%. 
The PDE and EEE are fitted according to Eq.~\ref{eq:doke} using the following electron recoil (ER) peaks, $^{131\rm m}$Xe (164\,keV), $^{129\rm m}$Xe (236\,keV), $^{127}$Xe (408\,keV), and $^{83\rm m}$Kr (41.5\,keV).
The PDE, EEE, and SEG$_{\rm b}$ in different datasets are summarized in Table~\ref{table:runConfig}.

The low energy calibration is carried out after set 5. The ER response is calibrated by injecting $^{220}$Rn into the detector. The $^{220}$Rn is produced by a foil $^{228}$Th source with an expected $^{220}$Rn emanation rate of 240\,Bq. The $^{220}$Rn rate observed in the detector is about 1.7\,Bq. In total, 1393 low-energy single-scatter ER events are collected within an $S1$ range from 2 to 135\,PE in the FV. The distribution of the ER calibration events in $\log_{10}(n_{\rm e}/S1)$ vs $S1$ can be found in Fig.~\ref{fig:calib_band}, where $n_{\rm e}$ is defined as $S2_{\rm b}/(\rm{EEE\times SEG_{\rm b}})$.

\begin{figure}[htbp]
    \centering
    \includegraphics[width=0.45\textwidth]{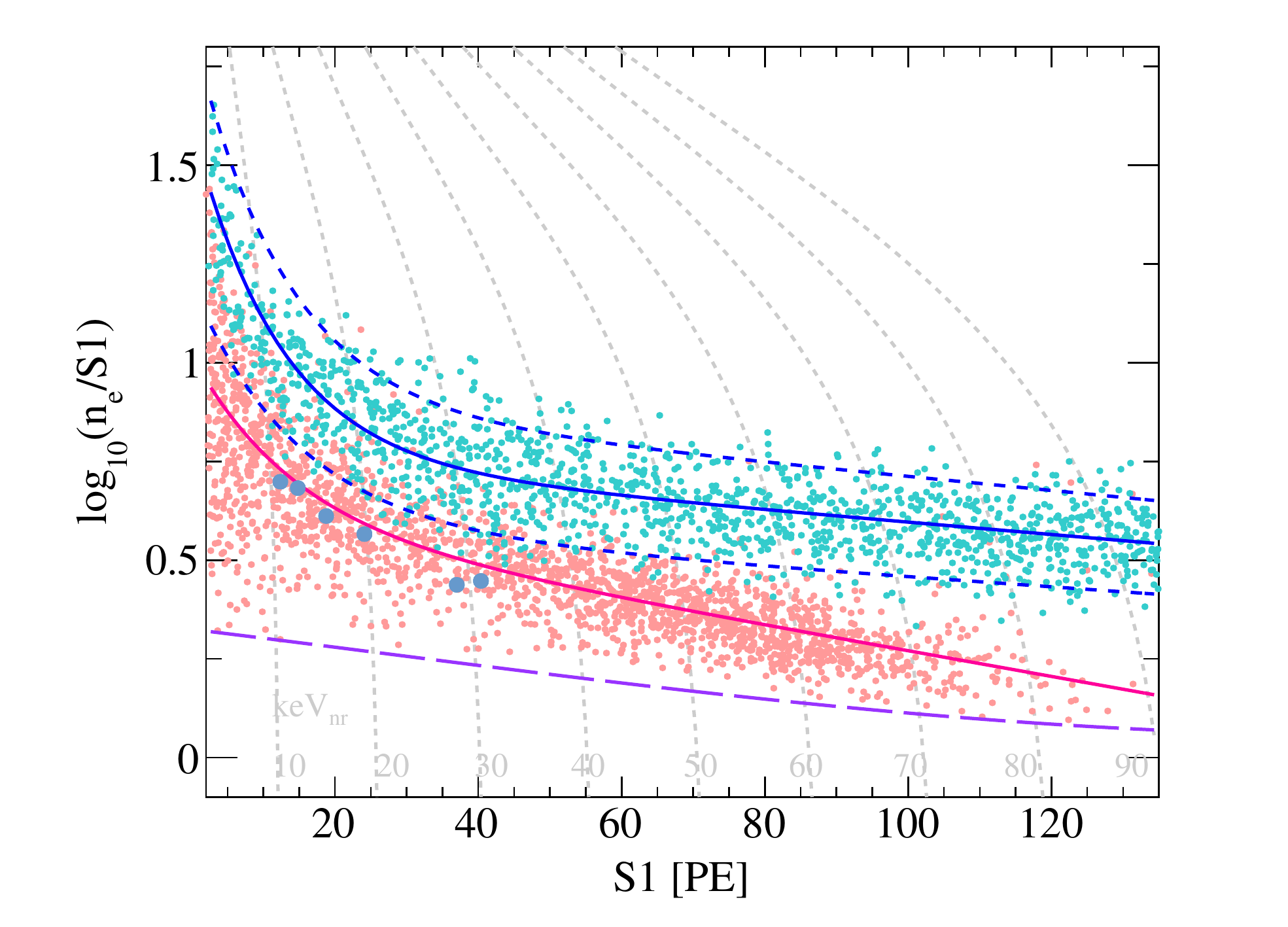}
    \caption{
    The distributions of $^{220}$Rn (cyan line) and D-D (magenta line) calibration events in $\log_{10}(n_{\rm e}/S1)$ vs. $S1$. The solid blue and red lines represent the fitted ER and NR medians, respectively, and the dashed blue lines are the corresponding 95\% quantiles of ER events. The D-D neutrons are selected within 120-520\,$\mu$s in drift time to avoid the so-called neutron-X events with partial energy deposition below the cathode. The six ER events from $^{220}$Rn calibration data located below the NR median line are highlighted.
    The dashed violet line represents the 99.5\% NR acceptance cut. The nuclear recoil energy in $\rm keV_{nr}$ is indicated with the gray dashed lines.
    }
    \label{fig:calib_band}
\end{figure}

The NR response is calibrated with two different neutron sources. The $^{241}$Am-Be source is deployed through three external horizontal tubes outside the inner cryostat at three different heights of the TPC. The deuteron-deuteron (D-D) neutrons are collimated horizontally via a beam pipe intruded into the water tank pointing to the center of the TPC. The distribution of the single-scatter D-D NR events is overlaid in Fig.~\ref{fig:calib_band}.

Since the data selection is made in $S1$ and $S2$, the total efficiency as a function of energy includes two major components, the signal reconstruction and detection efficiency, and the data quality cut efficiency. The signal reconstruction and detection efficiency is the ability for the readout and off-line software to correctly identify a true $S1$/$S2$, including the readout threshold (so-called BLS nonlinearity in Ref.~\cite{finalcpc}), signal clustering efficiency (15-sample), and the $S1$/$S2$ classification efficiency, all determined using data-driven methods. It also takes into account the event loss due to the requirements on number of hits ($\ge 2$), and selection ranges of $S1$ and $S2$, determined using the signal model simulation.
The efficiency of the data quality cuts
is determined using the calibration data by taking the calibration events within the $5\% - 95\%$ quantiles in Fig.~\ref{fig:calib_band}, and calculating the ratio of number of events with all cuts applied and that with all-but-this cut for the three classes of cuts described above.
The efficiencies separately determined from $^{220}$Rn, AmBe, and D-D calibration data are all consistent.
The total efficiency vs. nuclear recoil energy is shown in Fig.~\ref{fig:efficiency}.

\begin{figure}[htbp]
    \centering
    \includegraphics[width=0.45\textwidth]{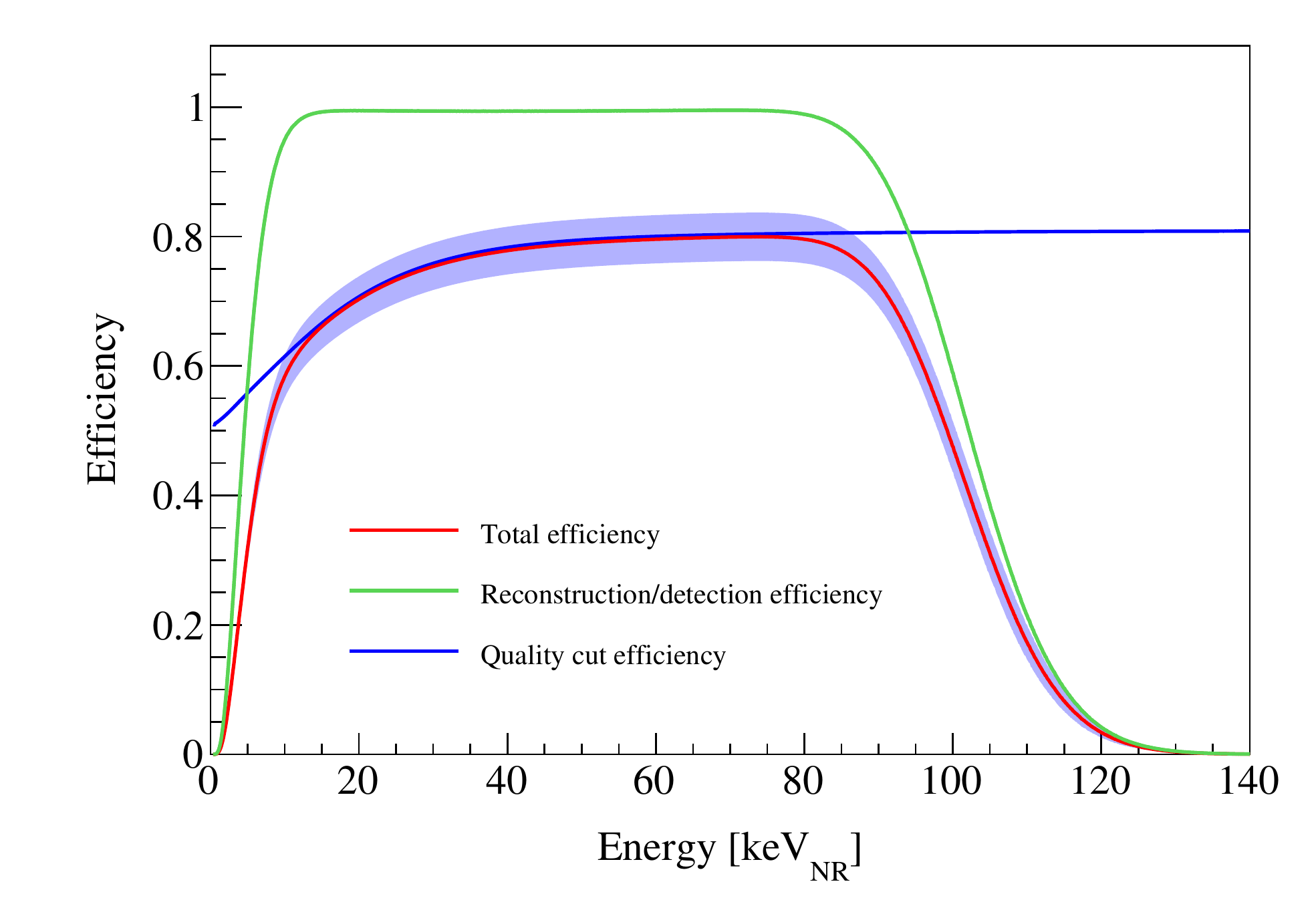}
    \caption{
    The reconstruction and detection efficiency (green line), data quality cut efficiency (blue line), and the total efficiency (red, with the shaded band representing the uncertainty) as a function of the nuclear recoil energy.
    }
    \label{fig:efficiency}
\end{figure}

A standard unbinned likelihood function is defined to perform a simultaneous fit of ER and NR response models based on the calibration data ($^{220}$Rn, $^{241}$Am-Be, and D-D) in ($S1$, $S2_{\rm b}$), with fast detector simulation including the following effects: photon detection, electron drifting, diffusion, extraction and amplification, nonuniformities of $S1$ and $S2_{\rm b}$, and detection efficiencies~\cite{PandaX-II:2021jmq,XENON:2019izt}. The ER and NR response models follow the standard NEST\,2.0 construction~\cite{NESTv2, szydagis2021review}, with the light yield, charge yield, and recombination parameters fitted. The likelihood function is minimized via Markov chain Monte Carlo~\cite{2013PASP..125..306F}, with fast detector simulation~\cite{PandaX-II:2021jmq,XENON:2019izt} boosted on GPUs. The best fit parameters are consistent with their nominal values in NEST\,2.0. The NR models obtained with AmBe or D-D data only are also in good agreement.
Binned-log-likelihood goodness-of-fit tests were made, with a resulting $p$-value of 0.38 and 0.78 for the ER and NR calibration data, respectively.
For sets 1-3, due to small differences in drift field in comparison to sets 4 and 5, the response model of sets 1-3 is extrapolated from that of sets 4 and 5 in accordance with the field dependence of light and charge yields in NEST\,2.0~\cite{NESTv2, szydagis2021review}, from which a less-than-1\% difference in the light yield is predicted. 
The same value of PDE is used for sets 1-3, but SEG$_{\rm b}$ and EEE (scaled from {\it in situ} ER peaks) are separately determined.

\begin{figure}[!htbp]
    \centering
    \begin{subfigure}{0.42\textwidth}
    \caption{$\log_{10}(n_e/S1)$ vs. $S1$}
    \includegraphics[width=0.95\textwidth]{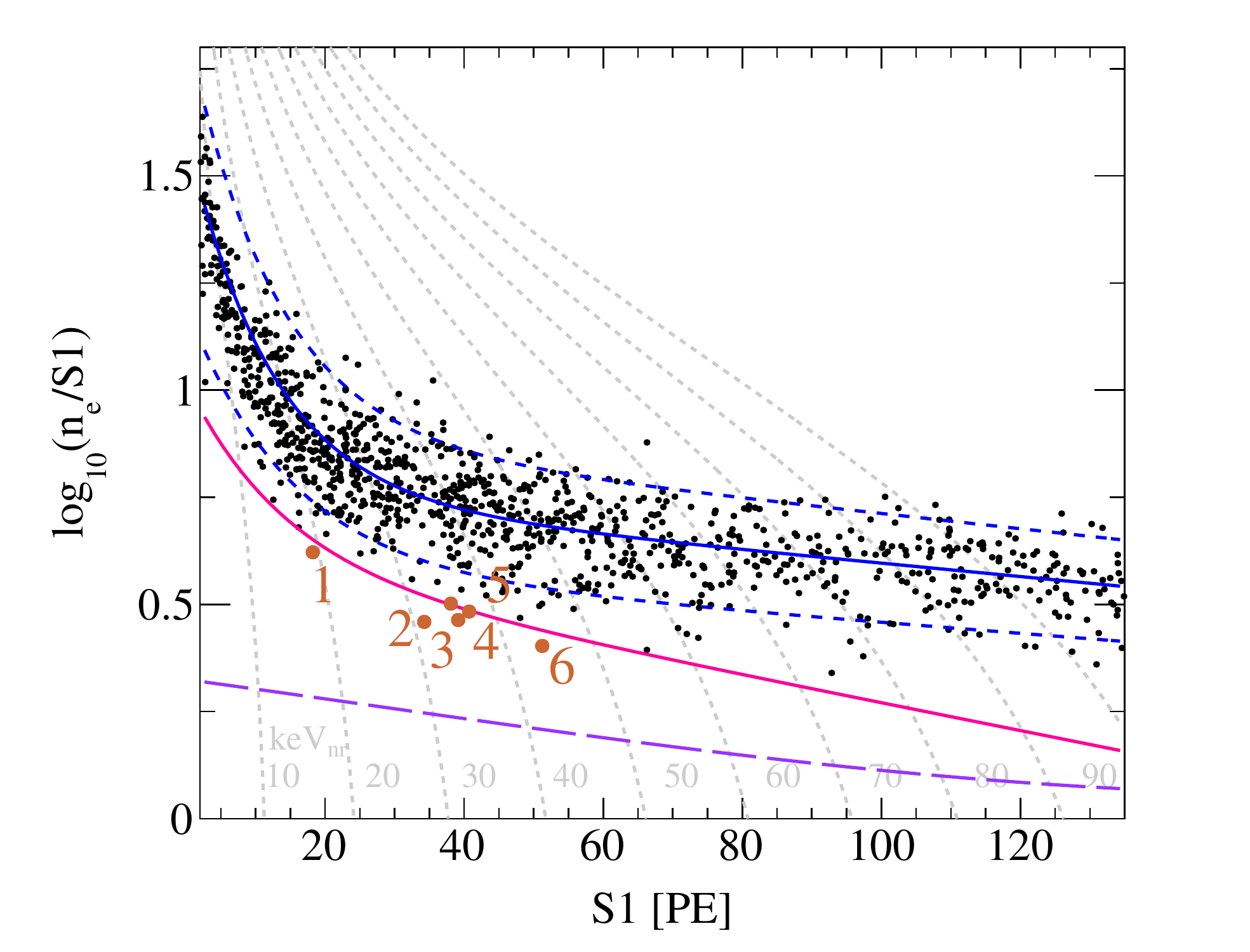}
    \label{fig:DM_band}
    \end{subfigure}
    \begin{subfigure}{0.42\textwidth}
    \caption{$z$ vs. $r^{2}$}
    \includegraphics[width=0.95\textwidth]{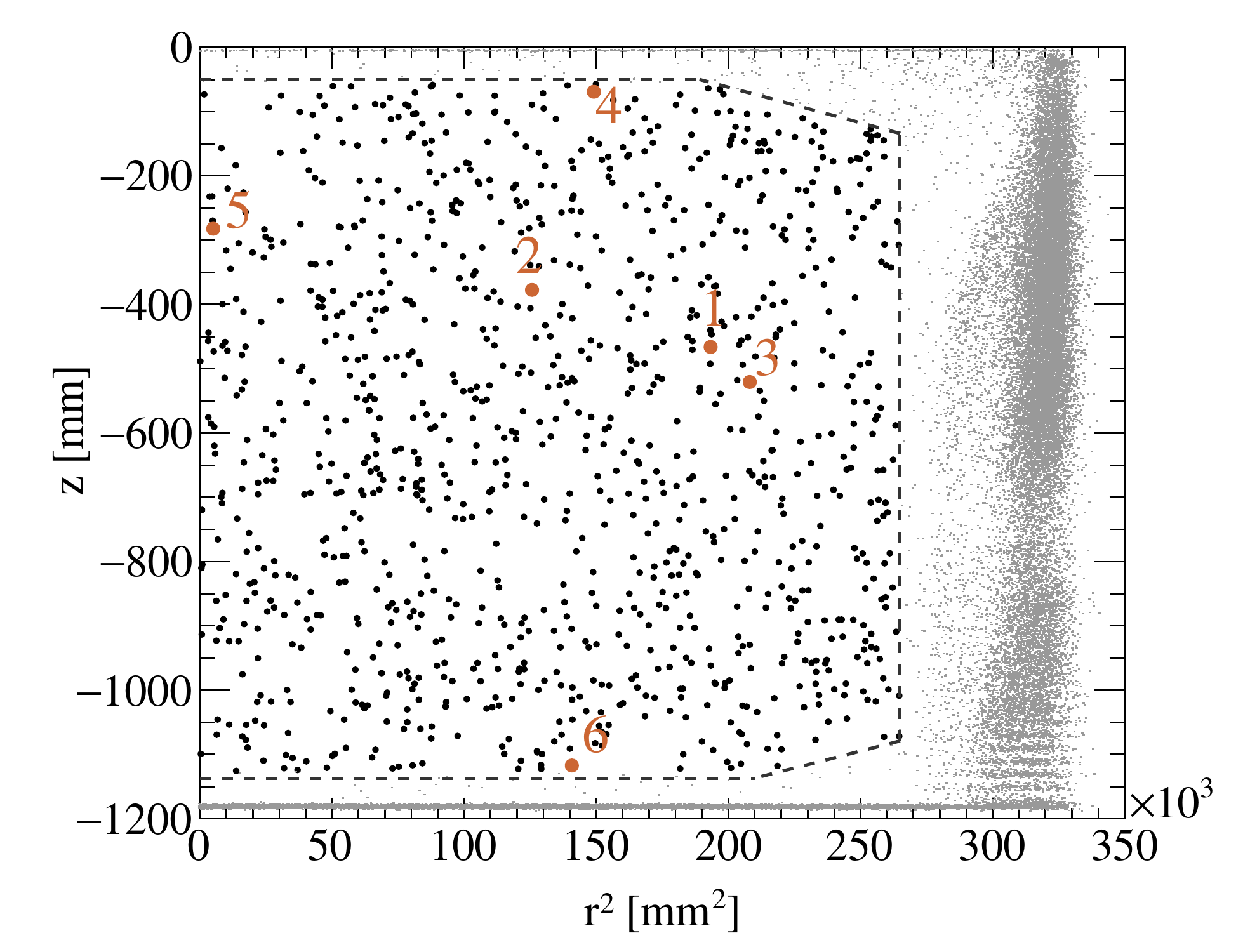}
    \label{fig:DM_vertex}
    \end{subfigure}
    \begin{subfigure}{0.42\textwidth}
    \caption{$y$ vs. $x$}
    \includegraphics[width=0.88\textwidth]{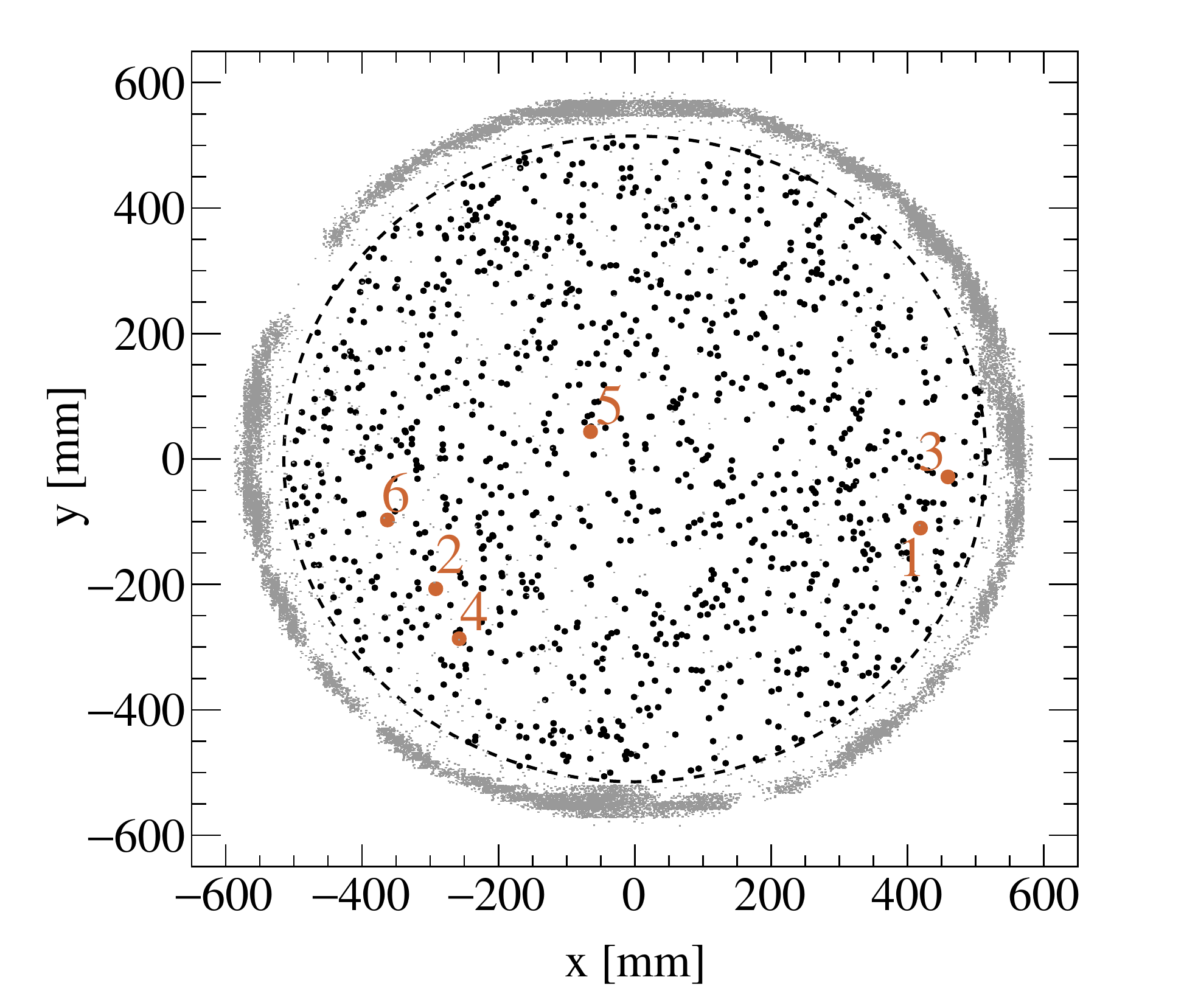}
    \label{fig:DM_pos}
    \end{subfigure}
    \caption{Distributions of the final dark matter candidates in $\log_{10}(n_e/S1)$ vs. $S1$ (a), $z$ vs. $r^2$ (b), and $y$ vs. $x$ (c). In (a), the solid blue and red lines are the ER and NR medians, respectively, and the dashed blue lines are the corresponding 95\% quantiles of ER events. The dashed violet line represents the 99.5\% NR acceptance cut.
    The nuclear recoil energy in $\rm keV_{nr}$ is indicated with the gray dashed lines. The six ER events located below the NR median line are highlighted in brown, with \#1 from set 3, \#2, \#4 and \#6 from set 4, \#3 from set 2, and \#5 from set 5. In (b) and (c), the dashed lines are projections of the FV, and black (light gray) dots represent events inside (outside). }
    \label{fig:DM_candidates}
\end{figure}

Aside from detector or data acquisition downtime, to eliminate stray electrons due to 
a previous energetic interaction, candidate events have to be separated by 22~ms from a previous event so that the contamination from leftover pulses can be neglected. This inserts a well-controlled deadtime of approximately 7.3\%. Data with abnormal isolated $S1$ rate, indicating excessive discharges from the electrodes and PMTs, are also removed from the analysis, reducing the live time by about 2.3\%. The resulting live time is 86.0-day. The dark matter candidates are selected using the following cut criteria. 
The ranges of $S1$ and $S2$ are $[2, 135]$ and $[80, 20000]$ PE, respectively. The veto PMT is required to see no coincidental photons during an $S1$.
The events are also required to be above the 99.5\% NR quantile (see Fig.~\ref{fig:calib_band}). The FV mass of about 2.67 tonne (with an uncertainty of 1.7\%), indicated in Fig.~\ref{fig:DM_vertex}, is determined based on the expected background distributions to optimize the sensitivity, with the material background from simulation, the internal contamination from data-driven estimate, and neutron background from a combined data-driven (rate) and simulation (vertex distribution) estimate.

The following major background components are considered in the dark matter analysis, with their rates summarized in Table~\ref{table:nominal}.

%

\begin{table*} [t]
\caption{\label{table:nominal}Expected background contributions to dark matter candidates for individual datasets. 
The "flat ER (data)" refers to a combination of radon, krypton, detector material background, solar neutrino, and $^{136}$Xe. For better statistical uncertainty, it is independently derived set by set from the data within the energy range from 18 to 30~keV, which is then used in the final likelihood fit.
The tritium values are obtained from unconstrained fit. The neutron, $^8$B, surface, and accidental background are assumed to be constant throughout the run.
The background-only best fit values and uncertainties are also shown, where the central values are used to generate pseudo datasets for the calculation of sensitivity. 
}
\centering
\begin{ruledtabular}
\begin{tabular}{cccccccccc}
                      & Set 1        & Set 2        & Set 3        & Set 4        & Set 5        & Total        & Below NR median & Best fit  \\ \hline
Rn                    & 6.9$\pm$3.8     & 42.8$\pm$23.5   & 22.7$\pm$12.5   & 162.0$\pm$88.9  & 112.1$\pm$61.5  & 346.5$\pm$190.2 & 1.42$\pm$0.78      &  -   &                    \\
Kr                    & 1.1$\pm$0.7     & 7.7$\pm$4.9     & 3.2$\pm$2.1     & 20.4$\pm$13.1   & 20.9$\pm$13.4   &  53.3$\pm$34.2   & 0.21$\pm$0.13      &  -   &                    \\
Material              & 0.8$\pm$0.1     & 5.7$\pm$0.7     & 2.4$\pm$0.4     & 15.2$\pm$1.9    & 15.6$\pm$1.9    & 39.7$\pm$5.0    & 0.16$\pm$0.02      &  -   &   \\
Solar $~\nu$ & 0.8$\pm$0.2 &5.4$\pm$1.1 &2.3$\pm$0.5 &14.3$\pm$2.9 &14.6$\pm$2.9 & 37.4$\pm$7.5&0.16$\pm$0.03 &- & \\
$^{136}$Xe & 0.7$\pm$0.1 &4.6$\pm$0.9 & 1.9$\pm$0.4 & 11.8$\pm$2.4 & 12.1$\pm$2.4 &31.1$\pm$6.2 & 0.05$\pm$0.01 & - &  
\\\hline
Flat ER (data) & 4.0$\pm$2.9     & 54.5$\pm$10.5   & 12.2$\pm$4.9    & 240.5$\pm$21.8  & 180.9$\pm$18.9  & 492.1$\pm$31.2  & 2.06$\pm$0.14      &  509.6$\pm$22.8   &                  
\\
CH$_3$T                  & 17$\pm$5           & 88$\pm$11          & 21$\pm$6   & 258$\pm$24  & 148$\pm$17  & 532$\pm$32  & 5.1$\pm$0.3          &   532$\pm$32  &                    \\
$^{127}$Xe                 & 0.19$\pm$0.04   & 1.08$\pm$0.25   & 0.96$\pm$0.22   & 3.99$\pm$0.92   & 1.91$\pm$0.44   &8.13$\pm$1.07   & 0.12$\pm$0.02  &  8.41$\pm$2.08   &                    \\
Neutron               & 0.02$\pm$0.01 & 0.15$\pm$0.08 & 0.07$\pm$0.03 & 0.45$\pm$0.22 & 0.46$\pm$0.23 & 1.15$\pm$0.57 & 0.69$\pm$0.35      &  0.82$\pm$0.41   &                    \\
$^8$B                    & 0.01$\pm$0.01 & 0.05$\pm$0.03 & 0.03$\pm$0.02 & 0.26$\pm$0.13 & 0.29$\pm$0.15 & 0.64$\pm$0.32 & 0.62$\pm$0.31     &  0.61$\pm$0.17   &                          \\
Surface               & 0.01$\pm$0.01 & 0.07$\pm$0.02 & 0.03$\pm$0.01 & 0.18$\pm$0.05 & 0.18$\pm$0.05 & 0.47$\pm$0.13 & 0.42$\pm$0.12      &  0.44$\pm$0.11   &                    \\
Accidental            & 0.04$\pm$0.01 & 0.32$\pm$0.05 & 0.03$\pm$0.01 & 0.99$\pm$0.18 & 1.05$\pm$0.21 & 2.43$\pm$0.47 & 0.80$\pm$0.15      &  2.31$\pm$0.45   &                    \\ 
Sum                &   21$\pm$6 &144$\pm$15  & 34$\pm$8 & 504$\pm$32 & 333$\pm$25      &   1037$\pm$45    &    9.8$\pm$0.6      &    1054$\pm$39   &\\\hline
Data                &     21      &     148      &     34      &     496      &     359      &     1058      &    6       & \\
\end{tabular}

\end{ruledtabular}
\end{table*}

The detector materials have been assayed by the high-purity germanium detector, and  the background due to material radioactivity is dominated by the PMTs and the stainless steel vessels~\cite{hongguang}. 
The energy and position distribution of the high-energy gammas in the data are consistent with expectations from simulation, and the integrated rates above 1~MeV agree within 14\%. The expected contribution to background in the dark matter window is $40\pm5$ events.

The radon background rate is measured {\it in situ} using alpha events. The decay of $^{222}$Rn is $4.2\pm0.1\,\mu$Bq/kg during sets 1, 2, 3, and 5, and $5.9\pm0.1\,\mu$Bq/kg during set 4 (increased due to radon emanation from the distillation tower during the online krypton distillation), and that from $^{220}$Rn is $0.07\pm0.01\,\mu$Bq/kg.
The expected low-energy radon background is dominated by $^{214}$Pb $\beta$s (decay lifetime $\sim$39\,min), which is not equally populated as their ancestors in the TPC as positive ions tend to drift toward and attach to the cathode~\cite{ma2020internal}. Its contribution is determined by taking the difference of the low-energy rates between sets 4 and 5. The overall contribution to the dark matter background is $347\pm190$ events.

The $^{85}$Kr $\beta$-decay background is estimated based on a correlated emission of $\beta$-$\gamma$ through the metastable state $^{85\rm m}$Rb (514~keV, 0.43\%). 
Assuming a $2\times10^{-11}$ isotopic concentration of $^{85}$Kr~\cite{Collon:2004xs}, a Kr/Xe ratio of $0.33 \pm 0.21$\,ppt is found. The expected background is $53\pm34$ events.

ER background due to solar neutrinos is estimated assuming the standard solar model, three-neutrino-flavor oscillation and the standard model anomalous magnetic moment~\cite{Billard:2013qya}. $^{136}$Xe two-neutrino beta decay is computed using the lifetime from Ref.~\cite{EXO-200:2013xfn}. The backgrounds from radon, krypton, detector materials, solar neutrino and $^{136}$Xe are combined into a ``flat ER (data)'' background in Table~\ref{table:nominal}, independently derived from data within the energy range from 18 to 30\,keV, and applied in the final dark matter fit.

Some number of tritium events are identified in the data. The origin is likely due to some leftover tritium from PandaX-II end-of-run calibration~\cite{PandaX-II:2020udv}. The event rate is allowed to float independently for each set, with a total fitted $532\pm32$~events in the FV and an average concentration of $5\times 10^{-24}$~mol/mol in xenon. The temporal variation in the data (Table~\ref{table:nominal}), particularly in between sets 4 and 5, indicates that gas circulation through hot getters may slowly reduce its concentration.

Some cosmogenically activated $^{127}$Xe is also identified in the data, which decays through electron captures. The background due to $L$-shell captures (5.2\,keV) is estimated based on the measured $K$-shell captures in the FV (33.2\,keV) and their expected ratio ($1:6$) to be $8\pm1$ events. Its decay (mean lifetime 52.5 days) is considered set by set in the final fit.

The long-lived progenies of radon attached onto the PTFE surface also contribute to the background, for example, through 
$^{210}$Pb $\beta$ decays. These events have a much suppressed $S2$ signal, likely due to the loss of electrons on the PTFE surface during the drift. The radial distributions of these events in different $S2_{\rm b}$ bins are obtained 
using $^{210}$Po surface events, tagged by S1s peaking around 30000\,PE (5.3\,MeV). The expected distribution in $S1$ and $S2_{\rm b}$ and the rate normalization is obtained using events reconstructed outside the PTFE wall but otherwise within the dark matter selection. The residual background in the FV is $0.5\pm0.1$ events.

The neutron background in the data is estimated using three methods. The first method is described in Ref.~\cite{hongguang}, but with updated radioactivities, selection efficiency and veto efficiency. The second method uses the single-scatter to multiscatter ratio of NR events. The third method follows the procedure in Ref.~\cite{Wang:2019opt}, with a predicted ratio between the 
single-scatter NR and high-energy neutron capture gammas. The residual neutron background in the dark matter data is $1.2\pm0.6$ events.

$^8$B neutrinos from the Sun can make coherent neutrino-nucleus scattering with xenon nucleus~\cite{ruppin2014complementarity}. 
This background is estimated to be 0.6$\pm$0.3 events.

The accidental background due to randomly paired $S1$ and $S2$ is studied by first identifying isolated $S1$ and $S2$ events, with a rate of 9.5~Hz ($S1$) and 0.0045~Hz ($S2$) and a standard deviation of 10.5\% ($S1$) and 12.7\% ($S2$), derived based on rates at different data-taking periods. The isolated $S1$s and $S2$s are randomly assembled in 
time, with selection cuts applied afterward. The remaining background in the dark matter sample is $2.4\pm0.5$ events, consistent with that obtained by selecting $S1$s and $S2$s from the data with a time separation beyond the maximum drift time.

Within the FV and dark matter selection window, 1058 final candidate events are identified. To take into account the difference in EEE and SEG$_{\rm b}$ between sets 1-2 and 3-5, we define $n_e = S2_{\rm b}/{\rm EEE}/{\rm SEG_b}$; thereby, the distribution of all events in $\log_{10}(n_e/S1)$ vs. $S1$ is shown in Fig.~\ref{fig:DM_band}.
Six events are identified below the NR median curve. Candidates are uniformly distributed in the FV, with position distributions in $z$ vs. $r^2$ and $y$ vs. $x$ displayed in Figs.~\ref{fig:DM_vertex} and \ref{fig:DM_pos}. 

\begin{figure}[htbp]
    \centering
    \includegraphics[width=0.45\textwidth]{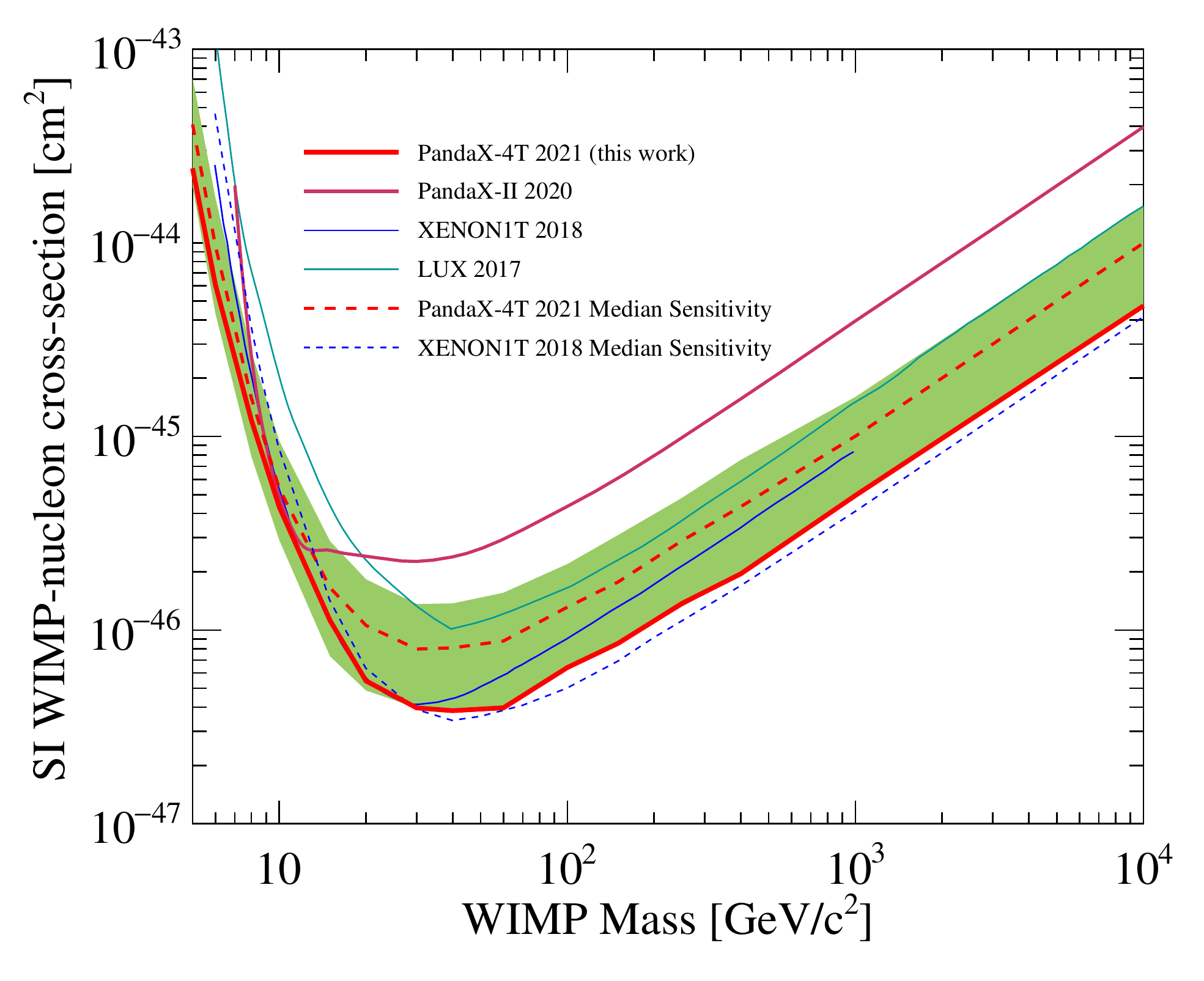}
    \caption{The 90\% C.L. upper limit from this work (unblind analysis) vs. $m_{\chi}$ for the SI WIMP-nucleon elastic cross section, overlaid with that from the full datasets of LUX 2017~\cite{Akerib:2016vxi_LUXNR}, XENON1T 2018~\cite{Aprile:2018dbl_xenon1tNR} and PandaX-II 2020~\cite{finalcpc}, obtained using blinded or salted analyses. The green band represents the $\pm1\sigma$ sensitivity band.
    The black and red dashed curves represent the median sensitivities of XENON1T and PandaX-4T, respectively.
    Results from XENON1T, LUX or PandaX-II partial datasets are not included in the figure. }
    \label{fig:limit}
\end{figure}

Dark matter signals are searched in our data using a profile likelihood ratio (PLR)
approach with a double-sided statistic construction~\cite{Baxter:2021pqo}. 
At each dark matter mass $m_{\chi}$ and its spin-independent (SI) elastic cross section with the nucleon $\sigma_{\chi,n}$, the NR rate and spectrum of 
the signal is computed using the recipe in Ref.~\cite{Baxter:2021pqo}. The probability density functions (PDFs) of the background and dark matter signals are both produced in $S1$ and $S2_{\rm b}$ using the aforementioned response models. 
A standard unbinned likelihood function is constructed~\cite{finalcpc}, with Gaussian penalty terms defined according to the uncertainty of the parameters in the response models and rates of individual background (Table~\ref{table:nominal}).
For our background-only fit, the goodness-of-fit $p$-value is 0.71. PLR scans are made on grids of ($m_{\chi}$, $\sigma_{\chi,n}$). No clear excess over background is observed. In Fig.~\ref{fig:limit}, the 90\% C.L. upper limit of SI cross section from our data is shown, together with $\pm1\sigma$ sensitivity band obtained from background-only pseudo data, as well as limits from previous experiments~\cite{Akerib:2016vxi_LUXNR, Aprile:2018dbl_xenon1tNR, finalcpc}. 
Our median sensitivity has improved from the PandaX-II final analysis~\cite{finalcpc} by 2.6 times at $m_{\chi}$ of 40\,GeV/$c^2$. Our limit is within the $\pm1\sigma$ sensitivity
band for $m_{\chi}$ below 25\,GeV/$c^2$, and goes slightly beyond $-1\sigma$ until about 250\,GeV/$c^2$, indicating a downward fluctuation of the background. The limit is, therefore, conservatively power-constrained to $-1\sigma$~\cite{Cowan:2011an}.
In comparison to XENON1T's final result, for $m_{\chi}$ below 20~GeV/$c^2$, our median sensitivity and exclusion limit are both stronger, which is primarily driven by our higher efficiency below 4 keV$_{\rm{nr}}$, attributed to the two-hit coincidence requirement. On the other hand, our median sensitivity is weaker than XENON1T for $m_{\chi}$ beyond 20~GeV/$c^2$, approaching a factor of 2.5 times or so for high-mass dark matter (DM). This is expected from the exposure (0.63 tonne-year vs. 1 tonne-year), efficiency difference at high recoil energy, and our higher background level due to tritium contamination.  More information can be found in the supplemental material.
Our new limit represents the most stringent constraint to DM-nucleon SI interactions, with the lowest excluded cross section value of $3.8\times10^{-47}$~cm$^2$ at $m_{\chi}$ of 40\,GeV/c$^2$.

In summary, we report the dark matter search results using the commissioning data from PandaX-4T, with a live exposure of 
0.63~tonne$\cdot$year. No dark matter candidates are identified above expected background. The strongest upper limit to date is set on the dark matter-nucleon spin-independent interactions, with the lowest excluded value of $3.8\times10^{-47}$\,cm$^2$ at 40\,GeV/c$^2$. PandaX-4T is undertaking a tritium removal campaign, after which normal physics data taking will start. The dark matter search sensitivity is expected to improve by another order of magnitude with a 6-tonne$\cdot$year exposure.


 
This project is supported in part by a grant from the Ministry of Science and Technology of
China (No. 2016YFA0400301), grants from National Science
Foundation of China (Nos. 12090060, 12005131, 11905128, 11925502, 11775141), 
and by Office of Science and
Technology, Shanghai Municipal Government (grant No. 18JC1410200). We thank supports from Double First Class Plan of
the Shanghai Jiao Tong University. We also thank the sponsorship from the
Chinese Academy of Sciences Center for Excellence in Particle
Physics (CCEPP), Hongwen Foundation in Hong Kong, and Tencent
Foundation in China. Finally, we thank the CJPL administration and
the Yalong River Hydropower Development Company Ltd. for
indispensable logistical support and other help.


\bibliography{pandax4t_wimp}
\section{Appendix}

\begin{figure}[h]
    \centering
    \begin{subfigure}{0.5\textwidth}
    \includegraphics[width=0.9\textwidth]{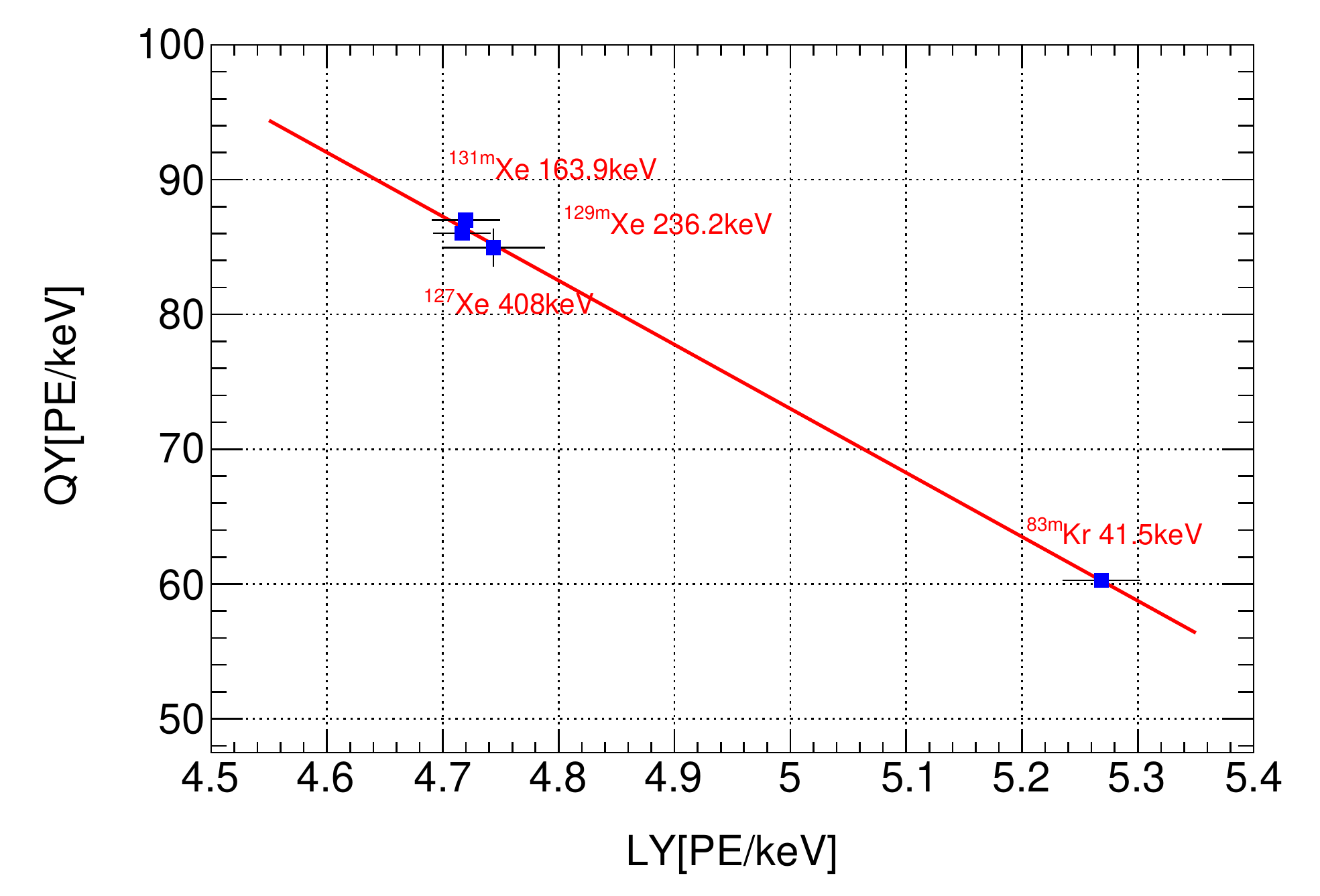}
    \caption{Charge yield vs. light yield in PE/keV and the linear fit.}
    \label{fig:Doke}
    \end{subfigure}
    \begin{subfigure}{0.5\textwidth}
    \includegraphics[width=0.9\textwidth]{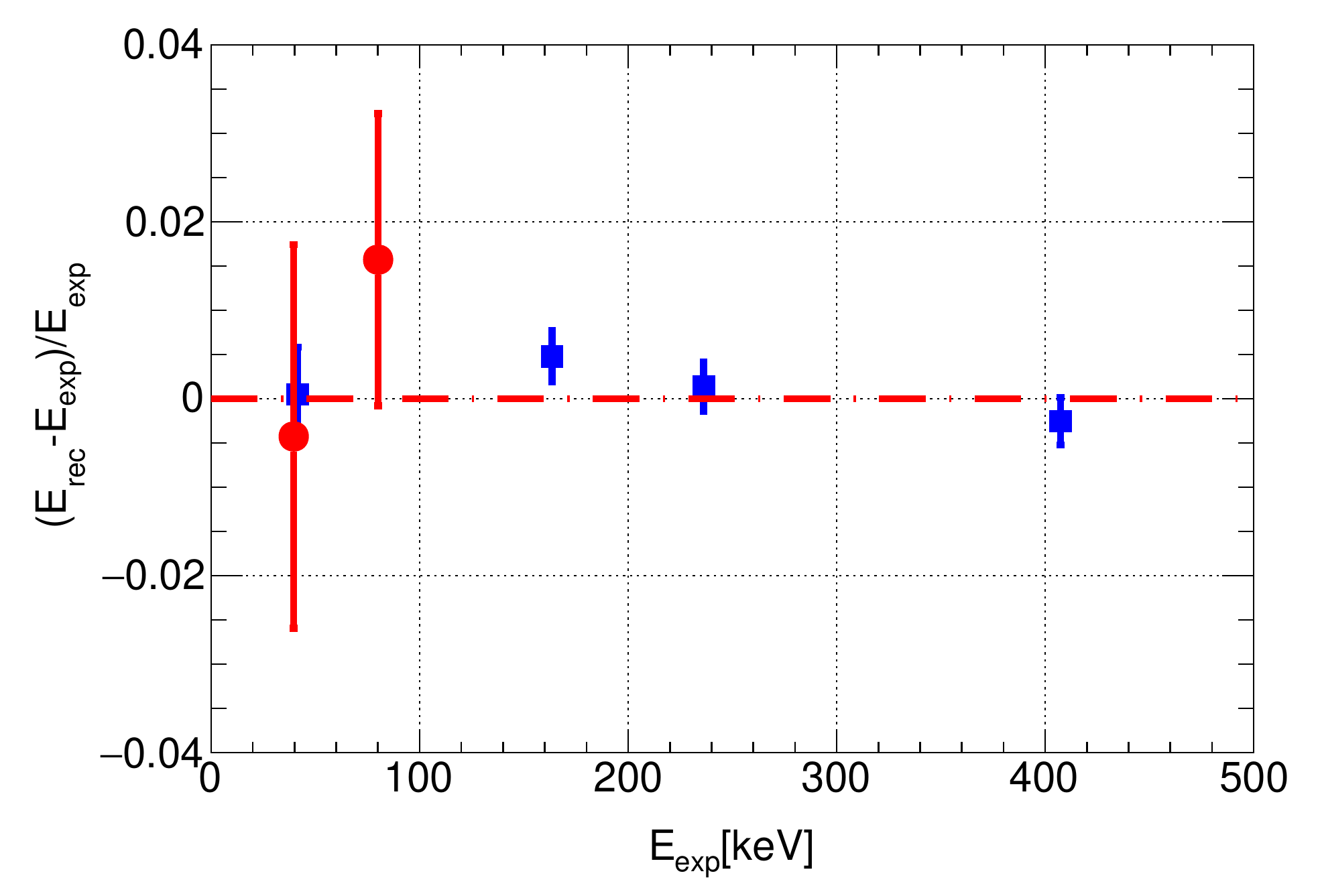}
    \caption{Fractional differences between the reconstructed and expected energy.}
    \label{fig:reconstruction}
        \end{subfigure}
    \caption{Charge yield vs. light yield (a) and energy reconstruction quality (b). Blue squares represent uniformly distributed electron recoil (ER) peaks, $^{131\rm m}$Xe (164~keV), $^{129\rm m}$Xe (236~keV), $^{127}$Xe (408~keV), and $^{83\rm m}$Kr (41.5~keV), which are used in the detector parameter fit (Eqn.1).
     Red points in (b) represent de-exitation gammas from $^{129}$Xe$^{*}$ (39.6~keV) and $^{131}$Xe$^{*}$ (80.2~keV) after neutron activation, with the NR components subtracted from the $S1$ and $S2$, which are not used in the linear fit. All uncertainties are dominated by systematic components.}
\end{figure}

\begin{figure}[h]
    \centering
    \includegraphics[width=0.45\textwidth]{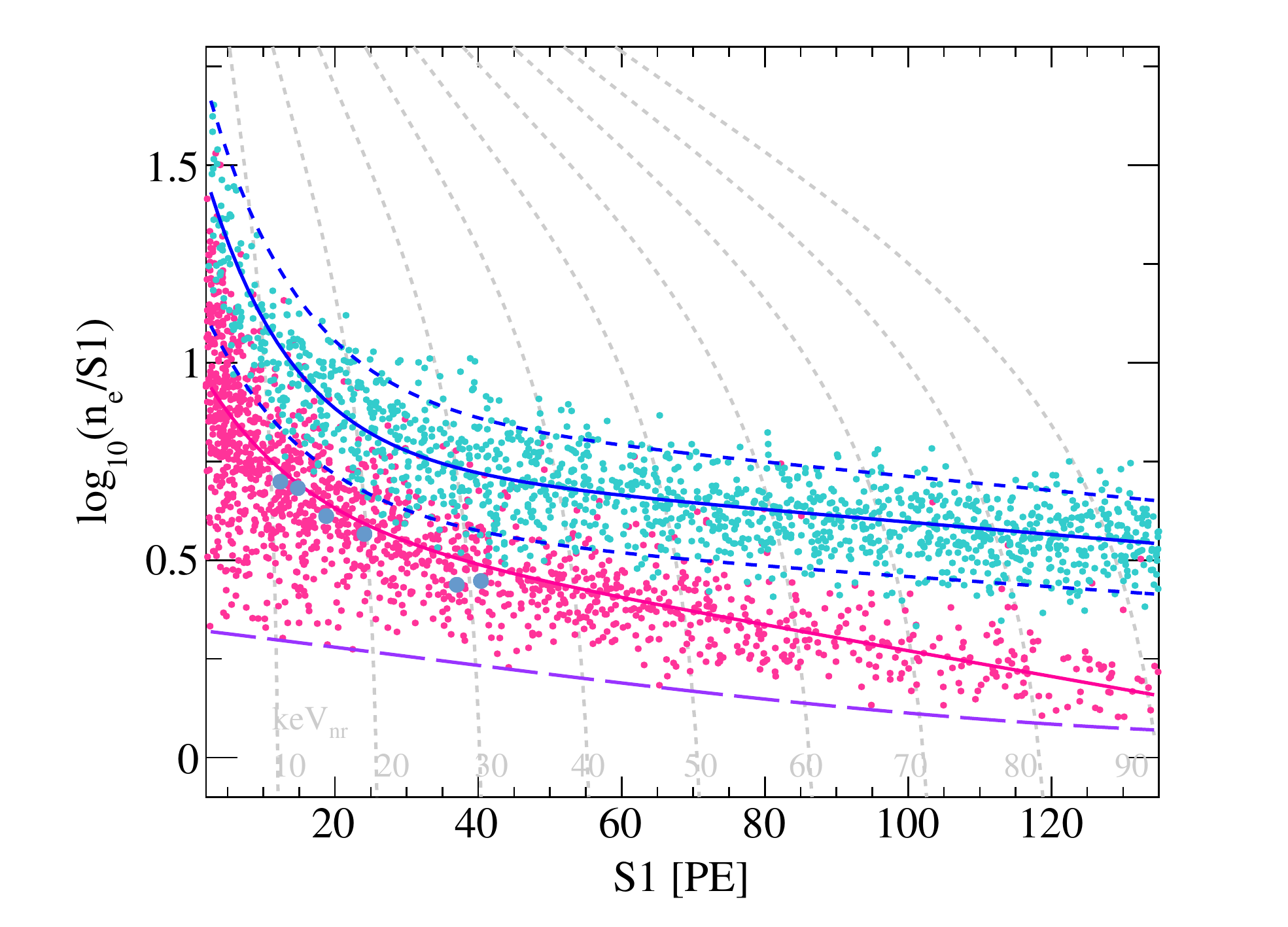}
    \caption{
    The distributions of $^{220}$Rn (cyan) and AmBe (magenta) calibration events in $\log_{10}(n_{\rm e}/S1)$ vs. $S1$. The solid blue and red lines represent the fitted ER and NR medians, respectively, and the dashed blue lines are the corresponding 95\% quantiles of ER events. The six ER events from $^{220}$Rn calibration data which are located below the NR median line are highlighted.
    The dashed violet line represents the 99.5\% NR acceptance cut. The nuclear recoil energy in $\rm keV_{nr}$ is indicated with the grey dashed lines.
    }
    \label{fig:calib_band}
\end{figure}

\begin{figure}[h]
    \centering
    \includegraphics[width=0.95\textwidth]{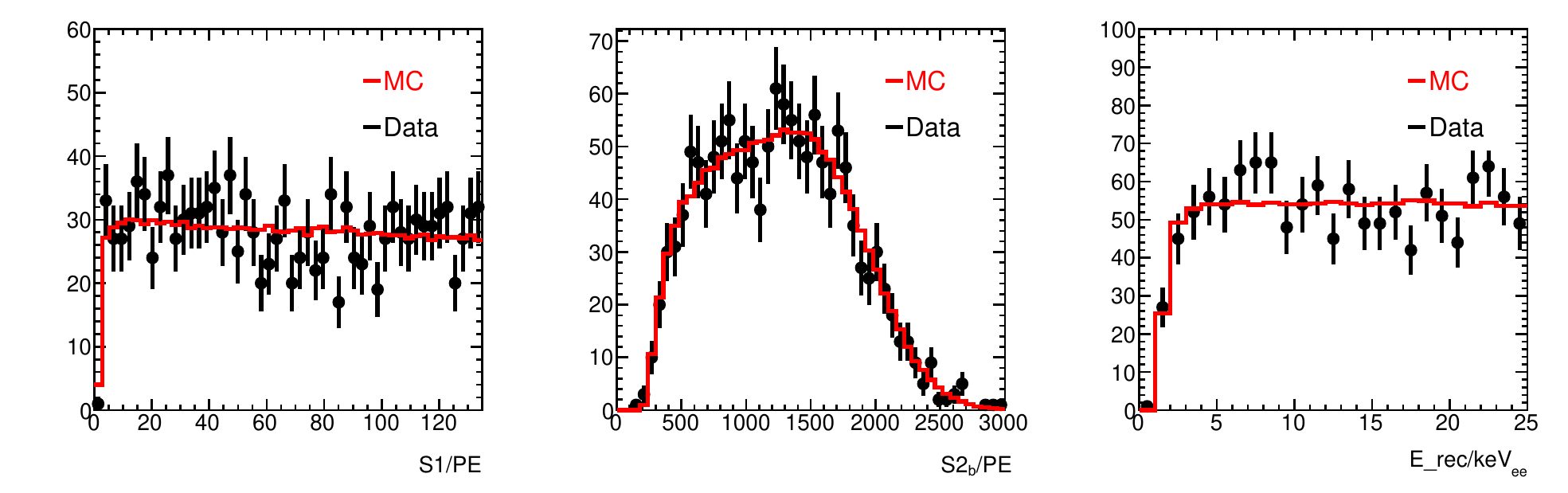}
    \caption{Distributions on S1, $\rm S2_{b}$ and reconstructed energy of Rn calibration data in comparison with the NEST-2.0-based Monte Carlo simulation. }
    \label{fig:Rn_MC_data}
\end{figure}

\begin{figure}[h]
    \centering
    \includegraphics[width=0.95\textwidth]{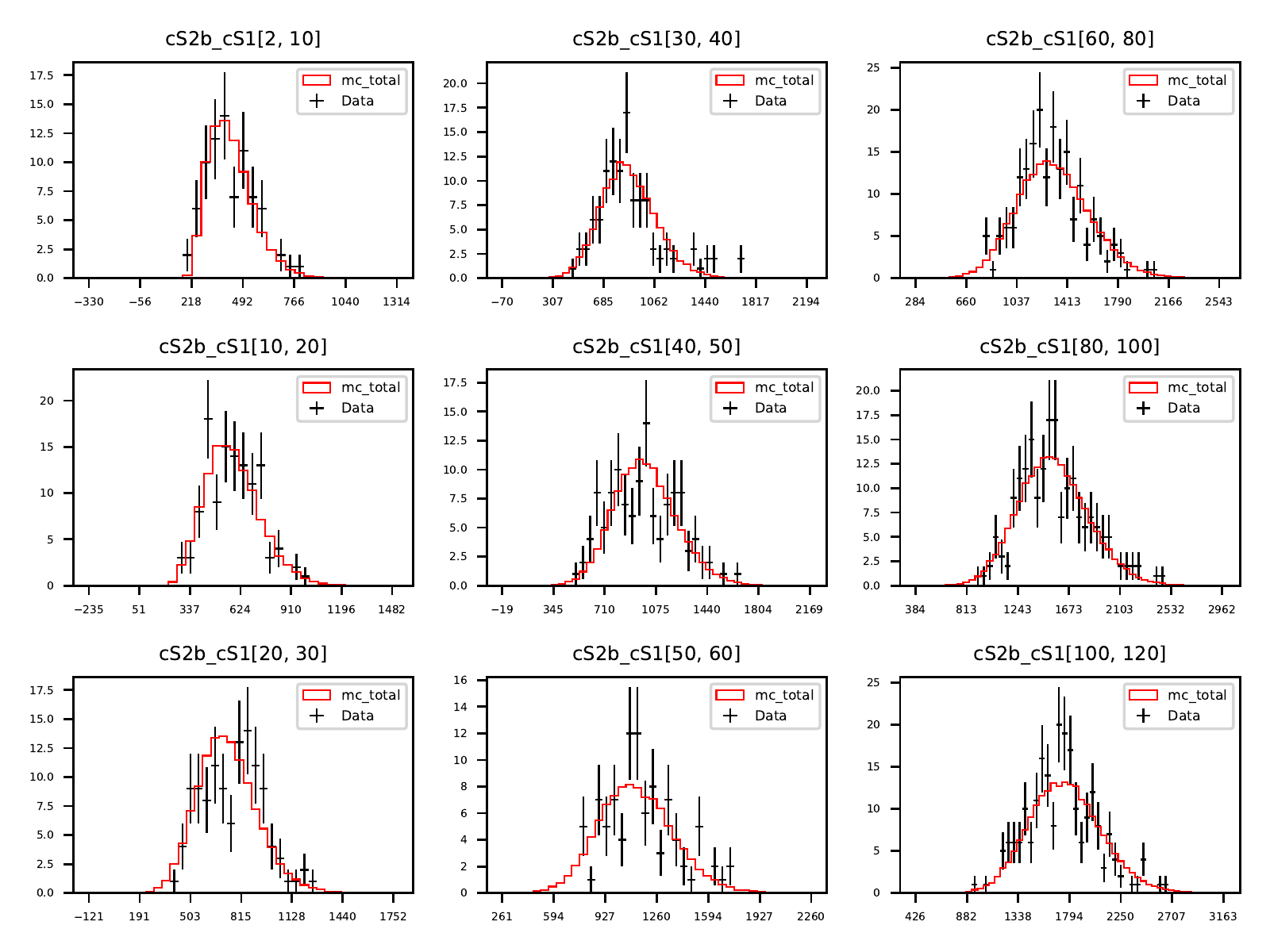}
    \caption{Distributions on $\rm S2_{b}$ with different S1 ranges of Rn calibration data in comparison with the NEST-2.0-based Monte Carlo simulation. }
    \label{fig:Rn_MC_data}
\end{figure}

\begin{figure}[h]
    \centering
    \includegraphics[width=0.95\textwidth]{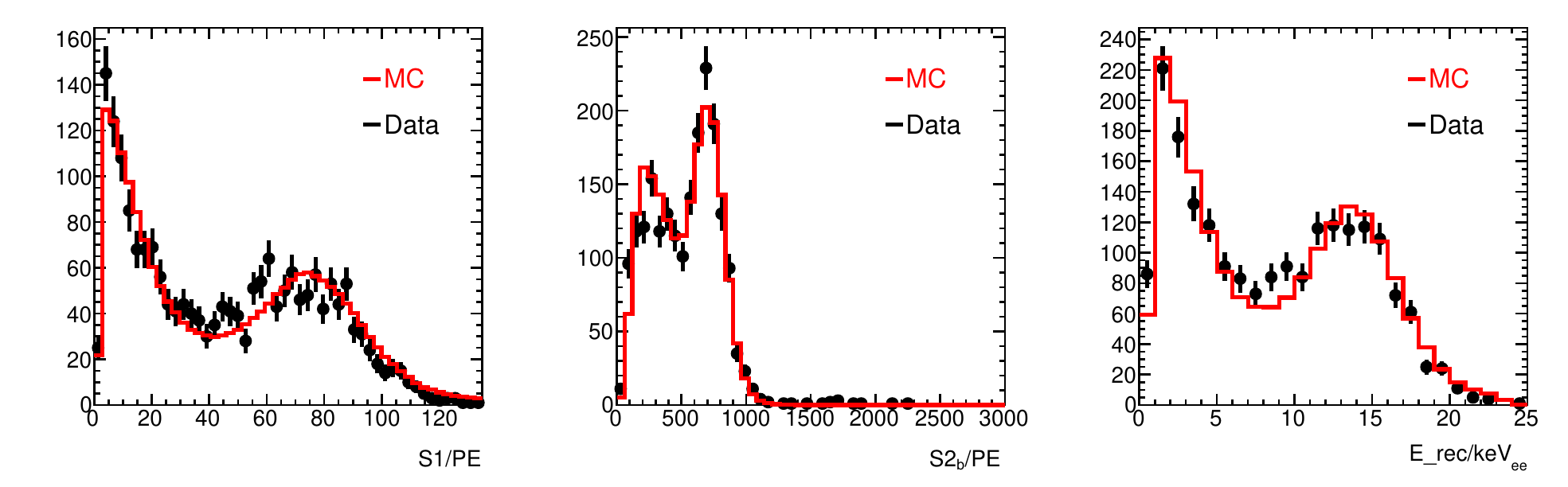}
    \caption{Distributions on S1, $\rm S2_{b}$ and reconstructed energy of DD calibration data in comparison with the NEST-2.0-based Monte Carlo simulation. }
    \label{fig:Rn_MC_data}
\end{figure}

\begin{figure}[h]
    \centering
    \includegraphics[width=0.45\textwidth]{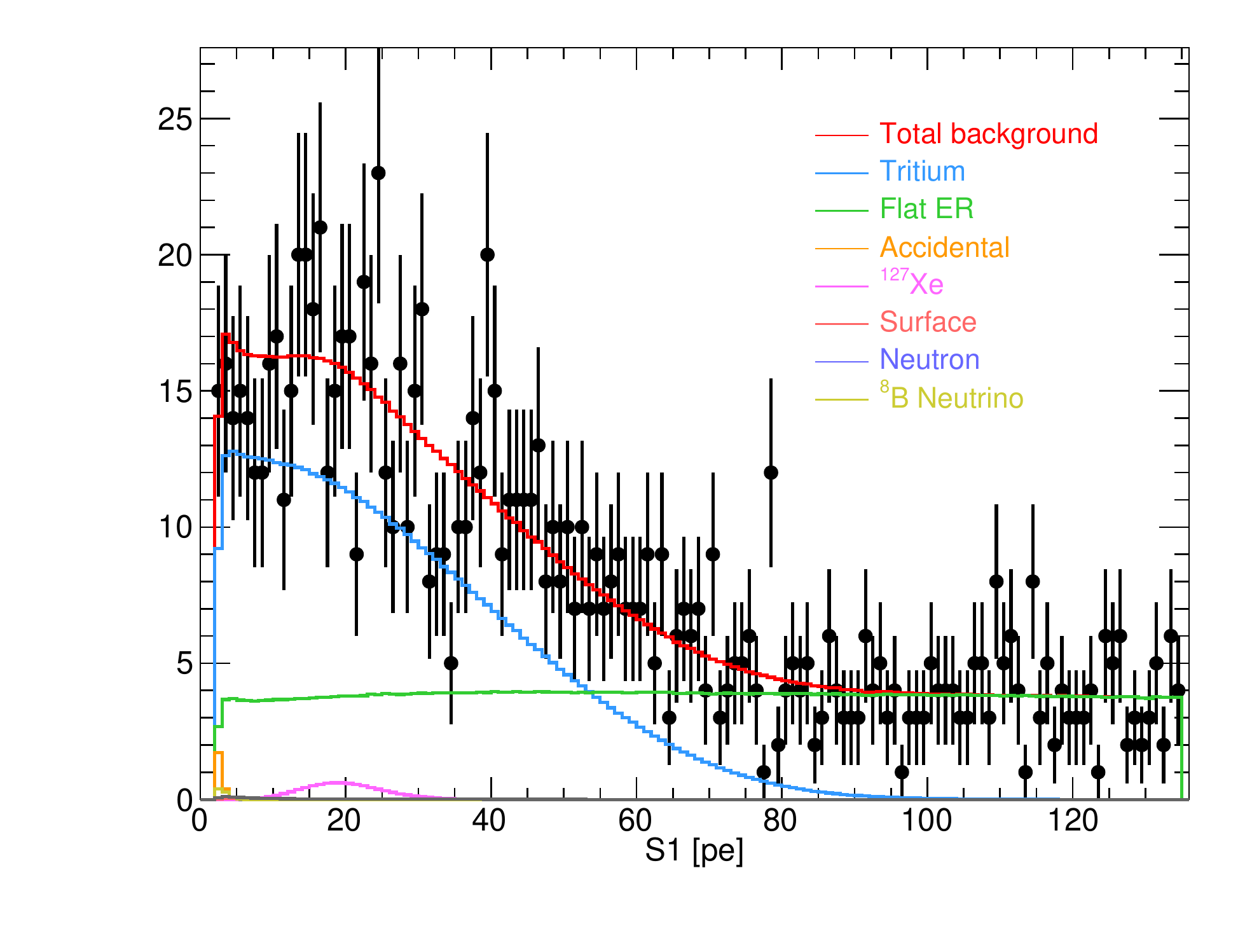}
    \caption{Distribution of dark matter candidates in $S1$ and background-only best fit.}
    \label{fig:BestFitting}
\end{figure}

\begin{figure}[h]
    \centering
    \includegraphics[width=0.45\textwidth]{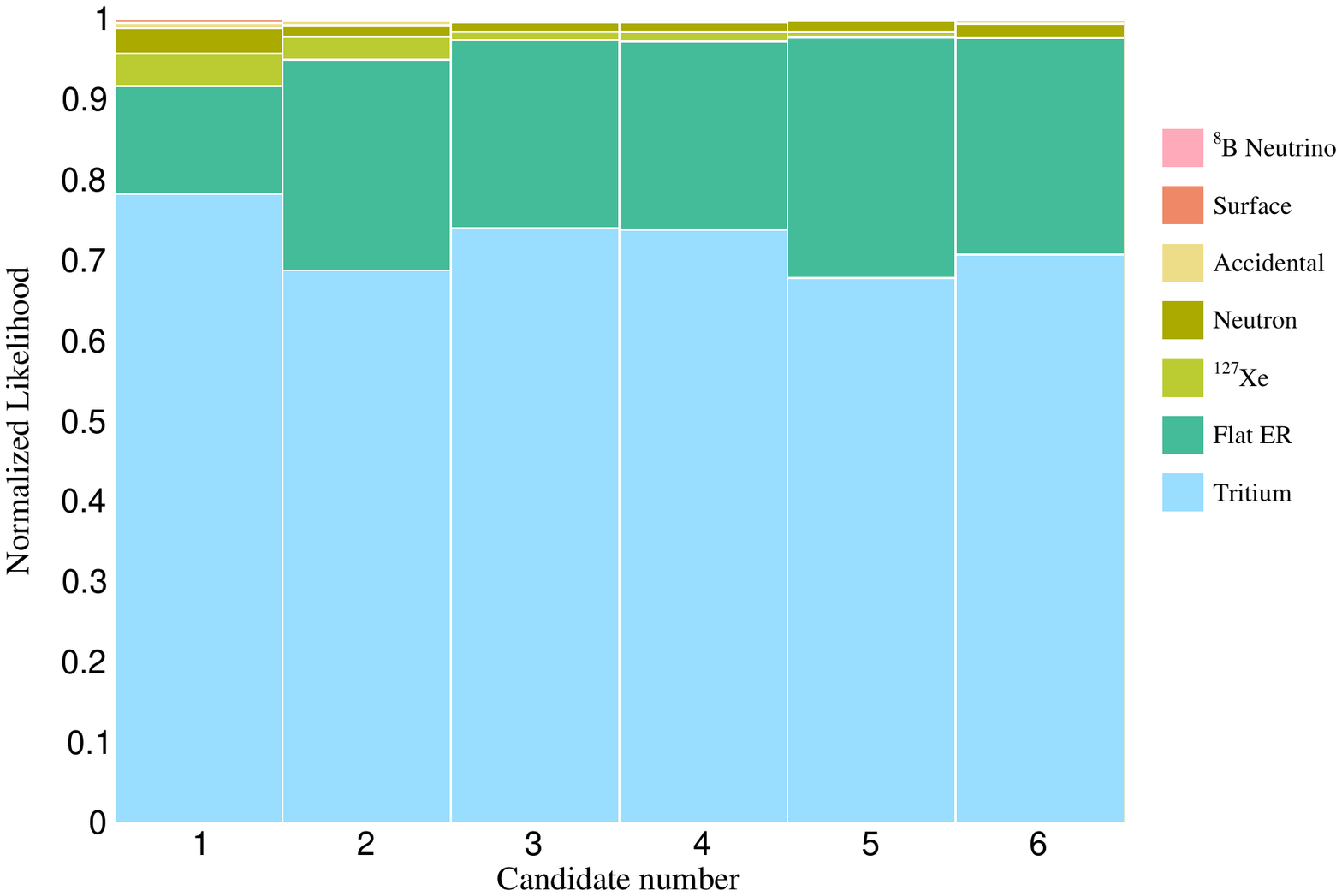}
    \caption{The best-fit component likelihood for the six events (see labels in Fig. 3 of the manuscript) located below the NR median line.}
    \label{fig:Likelihood}
\end{figure}

\begin{figure}[h]
    \centering
    \includegraphics[width=0.45\textwidth]{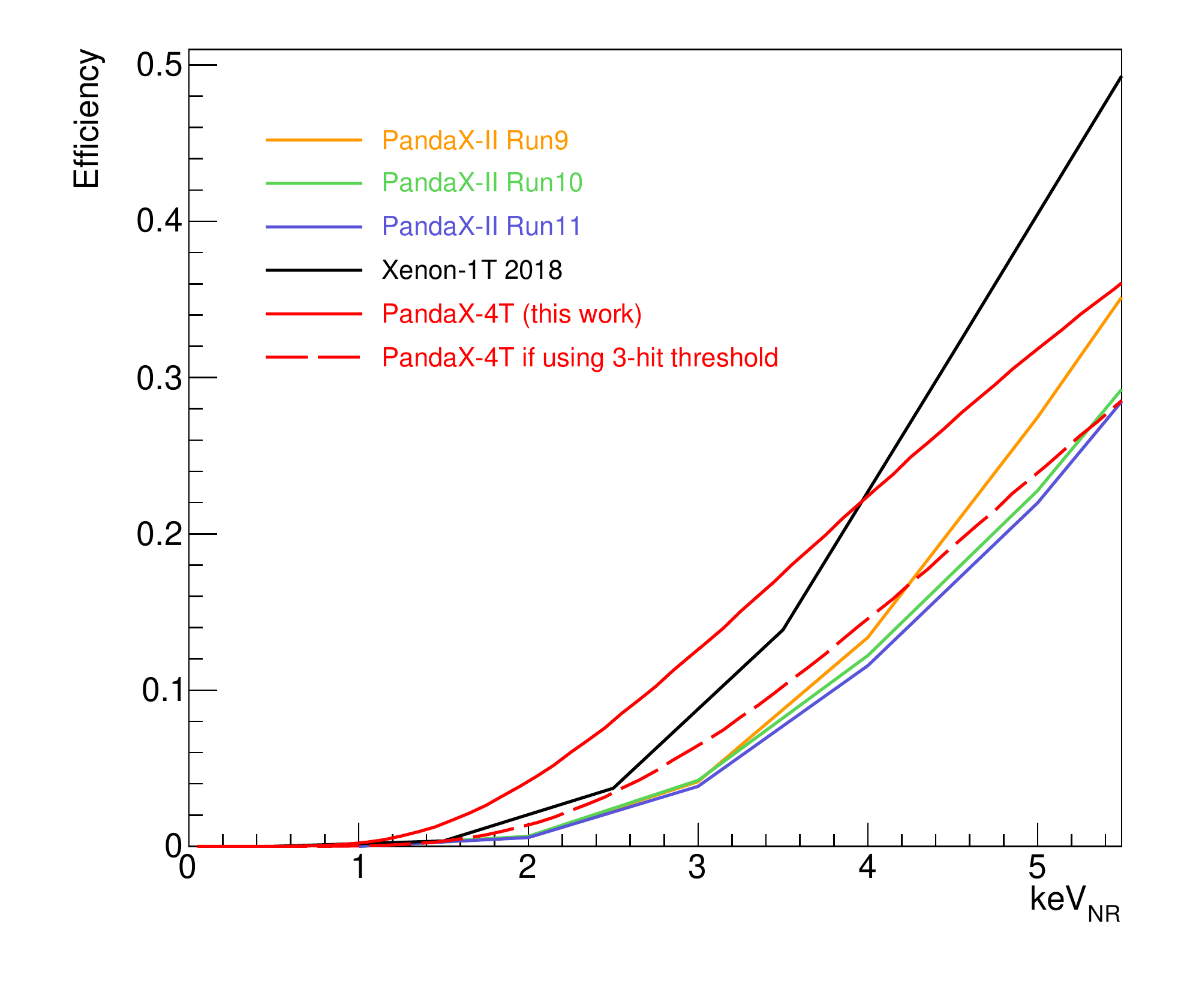}
    \caption{Comparison of total efficiencies of PandaX-4T with other experiments. Red dashed curve indicates the efficiency if with three-hit coincidence requirement on the S1 selection.}
    \label{fig:efficiency}
\end{figure}

\end{document}